\documentclass[pra, twocolumn, floatfix,nofootinbib, superscriptaddress]{revtex4-2}
\usepackage{dcolumn}
\usepackage{bm}
\usepackage{graphicx}
\usepackage{amsmath}
\usepackage{enumitem}
\usepackage{latexsym}
\usepackage{amsfonts}
\usepackage{amssymb}
\usepackage{array}
\usepackage{epsfig}
\usepackage[dvipsnames]{xcolor}
\usepackage{siunitx}
\DeclareSIUnit\hartree{Ha}
\DeclareSIUnit\angstrom{\text {Å}}

\usepackage{color}
\usepackage[colorlinks=true,bookmarks=false,linkcolor=RoyalBlue,urlcolor=RoyalBlue,citecolor=RoyalBlue,breaklinks]{hyperref}
\usepackage{physics}
\usepackage{bbold}

\begin{document}

\title{Mitigating the measurement overhead of ADAPT-VQE with optimised informationally complete generalised measurements}
\author{Anton Nyk\"anen}
\email{anton@algorithmiq.fi}
\author{Matteo A. C. Rossi}
\author{Elsi-Mari Borrelli}
\author{Sabrina Maniscalco}
\author{Guillermo García-Pérez}
\affiliation{Algorithmiq Ltd, Kanavakatu 3 C, FI-00160 Helsinki, Finland}

\date{\today}

\begin{abstract}
ADAPT-VQE stands out as a robust algorithm for constructing compact ans\"{a}tze for molecular simulation.
It enables to significantly reduce the circuit depth with respect to other methods, such as UCCSD, while achieving higher accuracy and not suffering from so-called barren plateaus that hinder the variational optimisation of many hardware-efficient ans\"{a}tze.
In its standard implementation, however, it introduces a considerable measurement overhead in the form of gradient evaluations trough estimations of many commutator operators.
In this work, we mitigate this measurement overhead by exploiting a recently introduced method for energy evaluation relying on Adaptive Informationally complete generalised Measurements (AIM).
Besides offering an efficient way to measure the energy itself, Informationally Complete (IC) measurement data can be reused to estimate all the commutators of the operators in the operator pool of ADAPT-VQE, using only classically efficient post-processing.
We present the AIM-ADAPT-VQE scheme in detail, and investigate its performance with ${\rm H_4}$ and ${\rm N_2}$ Hamiltonians and several operator pools.
Our numerical simulations indicate that the measurement data obtained to evaluate the energy can be reused to implement ADAPT-VQE with no additional measurement overhead for the systems considered here.
In addition, we show that, if the energy is measured within chemical precision, the CNOT count in the resulting circuits is close to the ideal one.
With scarce measurement data, AIM-ADAPT-VQE still converges to the ground state with high probability, albeit with an increased circuit depth in some cases.
\end{abstract}
\maketitle

\section{Introduction}
Quantum computing holds the potential to solve problems that are intractable on classical computers.
To unlock the full potential of the technology, error-corrected fault-tolerant quantum computing is needed, but even in the near-term, the continuously improving quantum computing hardware may be able to demonstrate practical computational advantage for certain important tasks \cite{preskill2018quantum,Barvyi2022}.
One of the fields in which near-term quantum computing hardware has the potential to provide advantage over classical state-of-the-art is the simulation of quantum systems themselves \cite{Daley2022}.
Especially, obtaining quantum advantage in the context of electronic structure problems in quantum chemistry \cite{kandala2017hardware,barkoutsos2018quantum,ollitrault2020quantum} and materials research \cite{lordi2021advances} would carry significant scientific and industrial impact. 

The general goal in quantum chemistry is to accurately estimate relevant properties of a molecular system described by a Hamiltonian $H_f$.
To determine the relevant chemical properties, the energy of the ground state of $H_f$ needs to be accurately estimated.
However, strong electronic correlations result in complex ground state wavefunctions that are difficult to describe with classical means.
For this reason, it was proposed that near-term quantum computers with the ability to prepare highly entangled states could be used to overcome these limitations \cite{Peruzzo2014}.
Within this context, the most commonly employed tool for the quantum simulation of molecular systems is the hybrid quantum-classical algorithm Variational Quantum Eigensolver (VQE), designed to find eigenvalues of quantum system Hamiltonians \cite{TILLY20221}.

In VQE, one considers a family of states that depend on some parameters and are encoded in a parameterised quantum circuit. The circuit should be shallow enough to survive the short coherence times of near-term quantum computers, and yet expressive enough to approximate the ground state for some value of its parameters \cite{McClean2016}.
For every value of the parameters, the resulting state represents the state of the electrons in a molecule and, as such, has an associated energy.
To find the ground state of the system, the parameters of the circuit are changed iteratively as to minimise the energy of the state.

VQE has been successfully applied to small electronic structure problems \cite{Peruzzo2014, Kandala2017, McArdle2020}, but there are several roadblocks that have so far prevented the scaling to practically relevant problem sizes \cite{Bittel_2021}.
Employing VQE for larger systems requires a robust methodology to design parameterised quantum circuits with the aforementioned properties.

Early attempts relied on so-called chemistry-inspired ans\"{a}tze, such as the Unitary Coupled Cluster Singles and Doubles (UCCSD) \cite{Peruzzo2014}, which can approximate the ground state relatively well.
However, these circuits are generally too deep to be executed within the coherence times of current devices, and UCCSD and its many variations have been shown to perform well only for systems with relatively low correlations \cite{Lee2019}, which can be solved classically efficiently.

In order to limit the circuit depth, it was later on proposed to use hardware-efficient ans\"{a}tze \cite{Kandala2017, Mitarai_2019}; parameterised circuits that, by construction, are shallow enough to be executed on hardware.
Yet, it has become apparent that only very limited accuracy can be achieved with these circuits. They typically exhibit landscapes of the energy cost function that are essentially flat almost everywhere in parameter space, rendering the optimisation unfeasible; a phenomenon better known as barren plateaus \cite{McClean2018, Zoe2021}.

The most promising approaches offering a solution to these problems are based on the Adaptive Derivative-Assembled Problem-Tailored Ansatz Variational Quantum Eigensolver (ADAPT-VQE) \cite{Grimsley2019}.
In ADAPT-VQE, the circuit is built iteratively such that, in each iteration, a new unitary operator is chosen from a pool and added to the ansatz based on its energy gradient.
Multiple variations of this algorithm have been studied, with different operator pools \cite{HoLun2021, Yordanov2021, Burton2022, VanDyke2022}, operator selection methods \cite{Burton2022, Lan2022, Anastasiou2022}, symmetry breaking and projections \cite{Bertels2022, Shkolnikov2021, Takashi2022}, and mutual-information assisted operator screening \cite{Zhang_2021}.

The adaptive ansatz construction both substantially limits the circuit depth and avoids the emergence of barren plateaus \cite{Grimsley2022}.
Moreover, ADAPT-VQE algorithms have been used for finding highly excited states \cite{ZhangFeng_2021} and quantum Gibbs thermal states \cite{Warren2022}.
However, it presents an important disadvantage: identifying the best operator to be added at each step introduces a drastic measurement cost, since the gradients of all the possible new operators must be estimated.
Several works address this measurement overhead \cite{Lan2022, Liu_2021, Sapova2022}, in some cases with significant cost reductions, although with the caveat that they result in deeper circuits.
Even in those cases, the overhead remains significant.

To use VQE for larger system sizes, besides the efficient construction of representative ans\"{a}tze, it is necessary to perform the energy estimation within a reasonable runtime. In the standard formulation, VQE algorithms require a prohibitively high number of measurements necessary for accurate energy estimation \cite{Gonthier_2022}.
In order to address this problem, many alternative measurement strategies have been proposed \cite{Jena2019,Yen2020,Huggins2021,gokhale2019minimizing,Crawford2021,paini2019approximate,bonet2020nearly,cotler2020quantum,hamamura2020efficient, kandala2017hardware, torlai2018neural,torlai2020precise,huang2020predicting,huang2021efficient,hadfield2022measurements, quek2021adaptive, miller2022hardware}. As a possible solution, some of the authors of this manuscript introduced an algorithm based on adaptive informationally complete measurements (AIM) \cite{Garc_a_P_rez_2021}.
The method exploits the informational completeness of generalised measurements to optimise them as to minimise the statistical error of the estimation of the energy.
In addition to reducing the total number of measurement rounds, also called shots, required to reach good accuracy in the estimation of the energy, the resulting IC data can be reused to estimate other expectation values using only classical post-processing, that is, without the need to perform additional measurements.

In this article, we propose to exploit the informational completeness of the measurement data to mitigate the measurement overhead of ADAPT-VQE.
More precisely, we consider the situation in which, at some iteration, the energy is measured using AIM in order to optimise the parameters in the current ansatz and the data is then reused to compute all the gradients in ADAPT-VQE.
While this is in principle possible by the nature of the IC data, we address the question of whether the precision of the gradient estimation is sufficient for this to perform in practice.
To that end, we introduce a framework to combine adaptive IC measurements with ADAPT-VQE, which we name AIM-ADAPT-VQE.
Our numerical simulations show that, with AIM-ADAPT-VQE, the measurement data obtained to determine the energy within chemical precision can be reused to implement ADAPT-VQE without increasing the CNOT counts for the systems analysed.
Interestingly, the ADAPT-VQE routine converges even if much less data is used, although sometimes with deeper circuits.

This article is organised as follows.
We first briefly review ADAPT-VQE and the AIM methodology in Sect.~\ref{sec:background}, and then introduce in detail the AIM-ADAPT-VQE algorithm in Sect.~\ref{sec:aim_adapt_vqe}.
In Sect.~\ref{sec:results}, we test the algorithm with ${\rm H_4}$ and ${\rm N_2}$ Hamiltonians and discuss the performance of the method for different measurement accuracies.
Section \ref{sec:conclusions} discusses some implications of the results and possible future work.

\section{Background}\label{sec:background}

In this section, we give a brief overview of the different algorithms that compose AIM-ADAPT-VQE.
We first summarise the general idea behind VQE, introduce ADAPT-VQE and some of its variants and, then, discuss the energy measurement cost problem along with the AIM method.

\subsection{VQE in a nutshell}

The general goal in quantum chemistry is to accurately estimate relevant properties of a molecular system described by a Hamiltonian 
\begin{equation}
H_f = \sum_{p, q} h_{p q} a_q^{\dagger} a_p+\frac{1}{2} \sum_{p, q, r, s} h_{p q r s} a_p^{\dagger} a_q^{\dagger} a_s a_r,
\end{equation}
where $a_p^{\dagger}$ and $a_p$ are the fermionic creation and annihilation operators, $ h_{p q}$ and $h_{p q r s}$ are one- and two-electron integrals, and $p,q,r$ and $s$ are spin-orbital indices.
As stated in the previous section, a Hamiltonian of this form can present highly complex ground states, which makes them hard to simulate for large system sizes, that is, when they involve a large number of spin-orbitals $N$.

In order to use VQE to tackle the problem, one first maps the fermionic problem into qubit space using so-called fermion-to-qubit mappings to construct a multi-qubit Hamiltonian $H$ with the same spectrum as $H_f$, so that the problem can be solved by a multi-qubit processor.
Many fermion-to-qubit mappings exist \cite{jordan_uber_1928, bravyi2002fermionic, Bravyi2017, jiang2020optimal, Chiew2022, Setia2018, Steudtner2018, Whitfield, Chien2020, Steudtner2019, miller2022bonsai}, and they generally lead to different qubit operators $H$ for the same fermionic Hamiltonian $H_f$, but the resulting operator can always be written as $H = \sum_k c_k S_k$, where $c_k \in \mathbb{C}$ and each $S_k \in \{ \mathbb{I}, X, Y, Z \}^{\otimes N}$ is a Pauli string.
The choice of mapping impacts several properties of the qubit operator $H$, which will be briefly discussed throughout the paper.

Once $H$ has been obtained, one then chooses a parametrised quantum circuit, a family of states $| \psi(\vec{\theta}) \rangle$ that depends on some parameters $\vec{\theta}$.
The circuit can be executed on a quantum computer, and its corresponding energy $E(\vec{\theta}) = \langle \psi(\vec{\theta}) | H | \psi(\vec{\theta}) \rangle$ can thus be evaluated through a sequence of physical measurements.
The energy $E(\vec{\theta})$ is then minimised with respect to parameters $\vec{\theta}$ until convergence.
According to the variational principle, $E(\vec{\theta}) \geq E_0$, where $E_0$ is the ground state energy, so the minimisation process brings the state $| \psi(\vec{\theta}) \rangle$ closer to the ground state.
In particular, if $E(\vec{\theta}_{\mathrm{opt}}) = E_0$, then $| \psi(\vec{\theta}_{\mathrm{opt}}) \rangle$ is the ground state of $H$.
If successfully implemented, the algorithm returns the ground state energy $E_0$, and further physical properties of the molecular system can be extracted from $| \psi(\vec{\theta}_{\mathrm{opt}}) \rangle$.

\subsection{ADAPT-VQE}

The key feature of all ADAPT-based strategies is that the circuit is built adaptively for the problem at hand, instead of having a fixed ansatz structure.
At iteration $n$, the ansatz construction step in ADAPT-VQE adds a unitary operator $e^{\theta_i P_i}$ to the current circuit $| \psi^{(n-1)} (\vec{\theta}^{(n-1)}) \rangle$, so that the resulting circuit prepares the states $| \psi^{(n)} (\vec{\theta}^{(n)})  \rangle = e^{\theta_i P_i} | \psi^{(n-1)} (\vec{\theta}^{(n-1)}) \rangle$, where $\vec{\theta}^{(n)}$ indicates the set of $n$ parameters in the ansatz at iteration $n$.
The operator $P_i$ is chosen from a pool of operators that is defined beforehand.
To determine which element should be chosen from the pool, one needs to perform a series of measurements for each of them, which results in the aforementioned overhead.
Once the operator is added, the new variational circuit,
\begin{equation}
    \label{eq:adapt_state}
    \ket{\psi^{(n)} (\vec{\theta}^{(n)})} = \prod\limits_{i = 1}^n e^{\theta_i P_i} \ket{\psi^{(0)}},
\end{equation}
where $| \psi^{(0)} \rangle$ is an initial reference state (usually the Hartree-Fock state), is variationally optimised with respect to the set of parameters $\vec{\theta}^{(n)}$.
Many aspects of the algorithm can be varied.
In the following, we will shortly describe some of the features and variants of ADAPT-VQE.

\textbf{Operator pool}. Defining a representative pool is very important, as it determines how well the circuit is able to explore the Hilbert space.
The pool in the original ADAPT algorithm follows chemical intuition and consists of fermionic single and double excitations \cite{Grimsley2019}.
Moreover, the operators are spin-complemented, which means that each operator excites both $\alpha$ and $\beta$ spin-orbitals (up and down spins) and thus conserves spin symmetry.
The fermionic pool used in this paper is the spin-dependent version with operators $\hat{\tau}_p^q = a_q^{\dagger} a_p-a_p^{\dagger} a_q$ and $\hat{\tau}_{r s}^{p q} = a_p^{\dagger} a_q^{\dagger} a_r a_s-a_s^{\dagger} a_r^{\dagger} a_q a_p$ as follows:
\begin{equation}
\label{eq:spin_pool}
\mathcal{P} =
\{ \tau_{p_{\alpha}}^{q_{\alpha}}, \tau_{p_{\beta}}^{q_{\beta}}, \tau_{r_{\alpha} s_{\alpha}}^{p_{\alpha} q_{\alpha}}, \tau_{r_{\beta} s_{\beta}}^{p_{\beta} q_{\beta}},
\tau_{r_{\alpha} s_{\beta}}^{p_{\alpha} q_{\beta}},
\tau_{r_{\beta} s_{\alpha}}^{p_{\beta} q_{\alpha}}\},
\end{equation}
where $\alpha$ and $\beta$ indices indicate the spin orientation of the spin-orbital.
This pool conserves $S_z$ and $N$ symmetries but can break the $S^2$ symmetry. 

To apply fermionic operators in a quantum circuit, they need to be mapped into operators in qubit space using fermion-to-qubit transformations.
The mappings considered in this paper are the Jordan-Wigner (JW) \cite{jordan_uber_1928}, Bravyi-Kitaev (BK) \cite{bravyi2002fermionic} and the JKMN \cite{jiang2020optimal} mappings.
Due to the structure of the JW mapping, the fermionic operators in the pool result in qubit operators involving Pauli strings with many $Z$ operators, the purpose of which is to keep track of the fermionic anti-symmetries.
It has been found, however, that removing these has a minimal effect on the performance.
The resulting operator pool is called Qubit-Excitation-Based (QEB) pool \cite{Yordanov2021}.
Note that this pool works only for the JW mapping and similar simplifications for other mappings are not straightforward.

A further simplification to the operators can be made by splitting the sums of the mapped fermionic operators (QEB operators with JW mapping), $\tau \mapsto \sum_k c_k S_k$, with $\tau \in \mathcal{P}$ and $S_k \in \{ \mathbb{I}, X, Y, Z \}^{\otimes N}$, into separate terms.
This creates the qubit-ADAPT pool, defined in qubits space in terms of the operators $P_k = i S_k$ \cite{HoLun2021}.
This pool simplifies the gate complexity but increases the number of variational parameters in the ansatz.

\textbf{Operator selection}. The operator selection can be done in many ways.
Ideally, the aim is to add gates that have maximum effect in taking the state towards the ground state of the system.
A heuristic to this end is based on the gradient of the energy with respect to the parameter in the unitary generated by operator $P_i$ when the parameter value is set to zero, which can be calculated by evaluating the commutator of $P_i$ with the Hamiltonian,
\begin{equation}
\label{eq:gradient_evaluation}
\begin{aligned}
\left.\frac{\partial E}{\partial \theta_i}\right|_{\theta_i=0} &=\left.\left[\frac{\partial}{\partial \theta_i}\left\langle\psi^{(n)}\left|e^{-\theta_i P_i} H e^{\theta_i P_i}\right| \psi^{(n)}\right\rangle\right]\right|_{\theta_i=0} \\
&=\left\langle\psi^{(n)}\left|\left[H, P_i\right]\right| \psi^{(n)}\right\rangle.
\end{aligned}
\end{equation}
Once the gradients for all the operators $P_i$ in the pool have been obtained, the operator with the highest absolute value is added to the ansatz.
Optionally, different strategies for gate ordering exist, when multiple gates are added in each step \cite{Lan2022, Anastasiou2022}.

Instead of maximum gradient, another possible selection criterion is the maximal energy reduction.
One can consider how much each operator decreases the energy by adding the gate to the current ansatz and optimising the whole circuit.
A more resource-efficient option is to only optimise the parameters of the trial operators.
In both cases, the operator or operators that lower the energy the most are added to the circuit.

\textbf{Optimisation.} There is further freedom in choosing the timing of the optimisations.
In the original ADAPT algorithm, the circuit is optimised after adding each gate.
To save resources, this can be relaxed and optimisation of the whole circuit can be done only between certain periods.
In order to advance the optimisation in the energy landscape and identify new operators to add, each newly added gate must be optimised individually.
This can be done efficiently with \emph{Rotosolve} \cite{Ostaszewski2019}, which finds the optimal parameter for the gate with few measurements based on the sine wave form of the energy curve as a function of the gate parameter.

\textbf{Measurement overhead.} As discussed previously, the measurement cost associated with the adaptive strategies is large, as each iteration of ADAPT requires information about the expectation values of many operators.
To analyse this in a more quantitative basis, notice that the number of operators in a pool, like the one in Eq.~\eqref{eq:spin_pool}, contains $O(N^4)$ operators and, for each of them, an expectation value of the form in Eq.~\eqref{eq:gradient_evaluation} must be measured.
Moreover, the naive evaluation of an expectation value of an $N$-qubit operator requires measuring each of the Pauli strings conforming it.
Thus, if we write the commutator in Eq.~\eqref{eq:gradient_evaluation} as a linear combination of Pauli strings, $\left[H, P_i\right] = \sum_k c_k S_k$, we need to measure the expectation value of each $S_k$ in the expansion
\begin{equation}
    \label{eq:naive_measurement}
    \left\langle\psi^{(n)}\left|\left[H, P_i\right]\right| \psi^{(n)}\right\rangle = \sum_k c_k \left\langle\psi^{(n)}\left| S_k \right| \psi^{(n)}\right\rangle.
\end{equation}
Since the Hamiltonian $H$ typically contains $O(N^4)$ Pauli strings itself, evaluating all the expectation values required in a single iteration of ADAPT-VQE can require measuring up to $O(N^8)$ Pauli strings.
The energy selection criterion discussed previously is of course much more demanding than the gradient based in terms of measurements, because of the need to optimise the circuit for each operator candidate.

Restricting the operator pool size and adding the operators according to different batching strategies have been suggested as possible ways to reduce the number of measurements in ADAPT-VQE \cite{HoLun2021, Lan2022, Liu_2021, Sapova2022}. Such approaches allow up to a quadratic improvement in the number of measurements, but  may result into deeper ansatz circuits if applied too greedily.

\subsection{Adaptive informationally complete measurements}

The simple measurement method introduced in the previous subsection is expensive in terms of shots.
Even if one is interested in estimating a single operator average, such as the energy of the system $\langle H \rangle$, the number of measurements required scales rapidly with the system size, especially for operators involving many Pauli strings.
Since any VQE-based algorithm requires measuring the energy repeatedly, the excessive number of shots has been pointed out as a potential roadblock for real-world applications \cite{Gonthier_2022}.
To overcome this issue, several ways to reduce the amount of measurements necessary to estimate an observable have been proposed \cite{Kandala2017, Izmaylov2020, Zhao2020, Huggins2021, Crawford2021}. 
Using adaptive IC generalised measurements to improve the measurement scheme on the fly has been introduced as one solution \cite{Garc_a_P_rez_2021}. 
The latter measurement algorithm, which we now review, was not only shown to reduce the energy estimation cost by orders of magnitude, but also to enable reusing the measurement data to evaluate other expectation values.
Importantly, multiple implementations of informationally complete positive operator-valued measures (IC POVM) have been proposed for gate-based quantum computers, either using ancillary qubits \cite{Garc_a_P_rez_2021}, exploiting higher energy levels of the physical qubits \cite{Fischer2022}, or by using randomised projective measurements \cite{Glos2022}.
In this paper, we consider only the class of POVMs used in Ref. \cite{Garc_a_P_rez_2021}.

For a single qubit $i$, these POVMs are generalised quantum measurements described by four so-called POVM effects, $\{ \Pi_{m_i} \}$.
That is, if the measurement is performed on a qubit in state $\rho^{(i)}$, the probability of obtaining outcome $m_i$ is $p_{m_i} = \mathrm{Tr}[\rho^{(i)} \Pi_{m_i}]$.
These POVM effects are positive operators $\Pi_{m_i} > 0, m_i=0, \dots, 3$ that add up to identity $\sum_{m_i} \Pi_{m_i} = \mathbb{1}$.
If they are linearly independent, and therefore span the space of linear operators in the Hilbert space of the qubit, the POVM is called informationally complete.
As such, any one qubit operator can be decomposed in the basis of the POVM effects.

Now, suppose every qubit in an $N$-qubit system is measured with a local POVM.
The probability to obtain an outcome $\mathbf{m} = (m_1, \dots, m_N)$, where $m_i \in \{ 0,1,2,3 \}$, given a state $\rho$, is $\mathrm{Tr}[ \rho \Pi_{\mathbf{m}}$], where $\Pi_{\mathbf{m}} = \bigotimes_{i=1}^N \Pi_{m_i}^{(i)}$.
If each of the single-qubit POVMs is IC, the set of $4^N$ effects $\{ \Pi_{\mathbf{m}} \}$ is IC in the space of linear operators of the $N$-qubit Hilbert space, $\mathcal{H}^{\otimes N}$.
In such case, any observable can be estimated with measurement data obtained from these POVMs, as we now show.

Since the effects $\{ \Pi_{\mathbf{m}} \}$ form a basis, an arbitrary observable $O$ can be decomposed as $O = \sum_{\mathbf{m}} \omega_{\mathbf{m}} \Pi_{\mathbf{m}}$, and its expectation value reads
\begin{equation}\label{eq:estimation}
\langle O \rangle = \mathrm{Tr}[\rho O] = \sum_{\mathbf{m}} \omega_{\mathbf{m}} \mathrm{Tr}[\rho \Pi_{\mathbf{m}}] = \sum_{\mathbf{m}} \omega_{\mathbf{m}} p_{\mathbf{m}},
\end{equation}
where $p_{\mathbf{m}}$ is the probability of obtaining outcome $\mathbf{m}$.
This allows the estimation to be obtained with a Monte Carlo approach: one can sample from the probability distribution $\{ p_{\mathbf{m}} \}$ with the quantum computer and average the $\omega_{\mathbf{m}}$ corresponding to each result.

The measurement is thus repeated $S$ times and measurement outcomes $\mathbf{m}_1, \dots , \mathbf{m}_S$ are obtained.
The expectation value of the operator can then be estimated as
\begin{equation}\label{eq:montecarlo}
\bar{O} = \frac{1}{S} \sum_{s=1}^N \omega_{\mathbf{m}_s}.
\end{equation}
In essence, any observable can be measured in this manner with the same measurement data.
Of course, the accuracy for the measurement will not be the same for all observables.
The standard error in the estimation is given by $\sigma_{O} = [(\langle \omega_{\mathbf{m}}^2 \rangle_{ \{ p_{\mathbf{m}} \} } - \langle \omega_{\mathbf{m}} \rangle_{ \{ p_{\mathbf{m}} \} }^2 ) / S ]^{1/2}$, which depends on the probability distribution $\{ p_{\mathbf{m}} \}$ and the operator decomposition. 
Regarding the latter point, the standard error depends on the so-called Pauli weight of the operator (the number of non-identity Pauli operators in the Pauli strings in its decomposition), and different fermion-to-qubit mappings lead to different Pauli weights for the same fermionic operator \cite{jiang2020optimal,Garc_a_P_rez_2021}.
Importantly, this approach enables producing accurate estimates of multiple observables with the same measurement data, even when the number of measurement shots $S$ is much smaller than $4^N$, the number of degrees of freedom in the states.
In fact, as shown in Ref.~\cite{Garc_a_P_rez_2021}, only a polynomial amount of measurements is required for typical few-electron operators, like the ones considered in this work, if an appropriate fermion-to-qubit mapping is used.

\begin{figure*}[t]
\includegraphics[width=.98\textwidth]{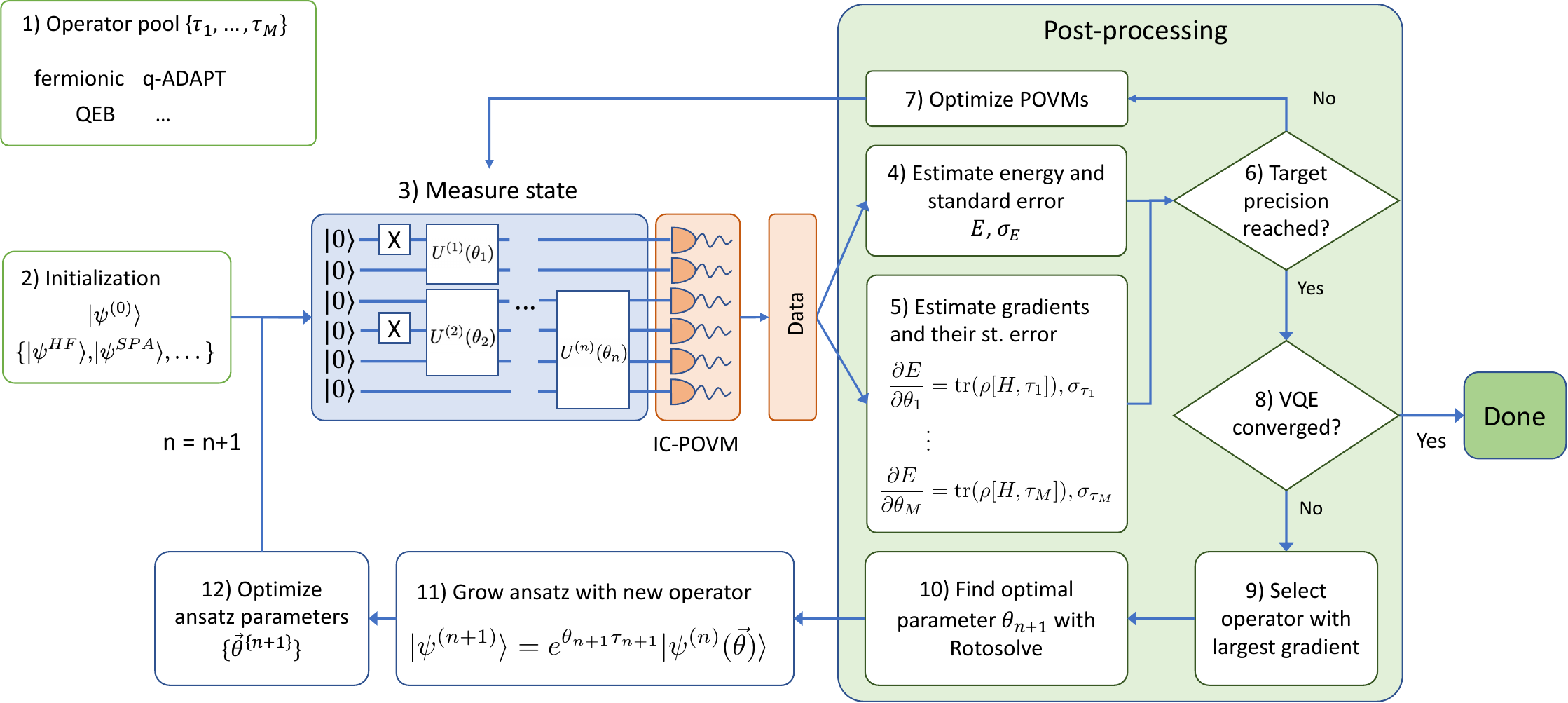}
\caption{Flowchart of the AIM-ADAPT-VQE algorithm. At each step of the ADAPT algorithm, the state is prepared on the quantum computer and measured with an IC POVM. A first batch of shots is collected and it is used to estimate the energy of the state, as well as the gradients of the operators in the pool. The POVM is optimised, a new batch of shots is collected and the process is repeated until the target precision is reached. The operator with the largest gradient is added to the ansatz, with an initial parameter chosen with \emph{Rotosolve}. All the parameters in the ansatz are then optimised using a classical optimiser. Notice that this step requires additional measurements. However, in this work, the latter task is performed using exact simulations (that is, without shot noise), as the focus is in the estimation of the gradients of the operators in the ADAPT pool.}
\label{flowchart}
\end{figure*}

To decrease the standard error for a given observable, the probability distribution should be modified, which can be achieved by modifying the POVM.
By parameterising the POVM, one can optimise over the parameter space to find a measurement that lowers the variance of the estimation, and hence becomes increasingly precise.
Moreover, this can be done by reusing the same measurement data that is used for the estimation of the observable.
In practice, a measurement procedure is split into iterations, where after each small batch of measurements, the data is used to adjust the POVM and estimate desired observables on-the-fly.
The process is repeated, and the number of shots in each round is updated according to a schedule, so that more measurements are used with more precise POVM \cite{Garc_a_P_rez_2021}.

\section{Estimating ADAPT-VQE commutators with IC data}\label{sec:aim_adapt_vqe}

In this section, we introduce AIM-ADAPT-VQE.
As discussed in the previous sections, the fact that IC measurement data can be reused to evaluate multiple expectation values of observables using classically efficient post-processing suggests a way to tackle the the overhead from the measurements of the gradients of many operators in each iteration of ADAPT-VQE.
In particular, if AIM are used in the optimisation of the parameters in the ansatz at each step (which would lead to a more efficient energy evaluation than with many other approaches), the IC data is automatically generated at no additional cost.
The idea is to then use Eq.~\eqref{eq:montecarlo} to estimate all the commutators required to implement the next step of ansatz construction in ADAPT-VQE.
In cases in which the ADAPT-VQE schedule does not require the optimisation of the current ansatz, some measurements will be needed, but we proceed in the same manner, that is, by optimising the AIM to measure the energy.

An important aspect of the routine is how to determine whether enough measurements have been conducted.
Typically, it is assumed that chemically precise ($\SI{1.6}{\milli\hartree}$) estimates of the energy are needed, since this is the level of precision required in real-world applications.
However, in the operator selection and at intermediate stages of the variational circuit optimisation, such precision might not be required.
In order to investigate this question, we propose two different strategies: energy-based and gradient-based target precision.
In the former, the AIM routine stops when the estimation of the energy is accurate enough.
In the latter, the criterion to stop is based on the relative error in the estimation of the ADAPT commutators.

The steps of the AIM-ADAPT-VQE algorithm with energy-based measurement target precision, as shown also in Fig.~\ref{flowchart}, are the following:
\begin{itemize}
    \item[1.] Prepare the operator pool and choose the fermion-to-qubit mapping.
    \item[2.] Initialise the ansatz. The most common choice is the Hartree-Fock state but other choices, like the Separable Pair Ansatz (SPA) \cite{Kottmann_2022}, may be used.
    \item[3.] Measure the state with an initial amount of shots.
    \item[4.] Estimate the energy and the standard error of the estimation.
    \item[5.] Estimate every commutator $[H, \tau_i]$ and their corresponding standard errors.
    \item[6.] Check if the standard error of the energy measurement is below the target threshold $T_E$. If so, skip to step 8.
    \item[7.] Optimise the POVM, update the measurement strategy according to the schedule, and go back to step 3.
    \item[8.] Check for convergence of the VQE routine based on the gradient norm of the operator pool or the variance of the energy, and stop the algorithm if it has converged.
    \item[9.] Choose the operator with the largest gradient.
    \item[10.] Find an optimal parameter for the new operator using \emph{Rotosolve}.
    \item[11.] Add the gate to the ansatz.
    \item[12.] If assigned in the schedule, optimise the whole circuit. Go to step 3.
\end{itemize}

In the gradient-based stopping version of the algorithm, to select the next operator, the goal is to be able to choose the one with the highest gradient while minimising the risk of choosing a wrong gate.
Often, many operators have gradients that are very close to each other and, in these cases, choosing any of these usually gives similar results.
The problem arises when the standard error of the estimation is close to or larger than the typical difference between gradients, as it is then likely for a wrong choice to be made.
In order to prevent these situations, we continue measuring until the relative error in the largest gradient is smaller than the target threshold. 
In this alternative measurement strategy, step six is changed to:
\begin{itemize}
    \item[6.] Check if the relative error in the estimation of the commutator with the current highest mean is below the target threshold $T_\tau$. If so, skip to step 8.
\end{itemize}

The measurements in the AIM-ADAPT-VQE algorithm are performed while the POVM are continuously optimised, using the same shot schedule as in Ref.~\cite{Garc_a_P_rez_2021}.
The performance of this optimisation is evaluated after each small batch of measurements by considering the standard deviation of the energy estimation, defined as $\epsilon_E = [ \langle \omega_{\mathbf{m}}^2 \rangle_{ \{ p_{\mathbf{m}} \} } - \langle \omega_{\mathbf{m}} \rangle_{ \{ p_{\mathbf{m}} \} }^2 ]^{1/2}$, where $\omega_{\mathbf{m}}$ are defined through the relation $H = \sum_\mathbf{m} \omega_{\mathbf{m}} \Pi_\mathbf{m}$, and $\{ \Pi_\mathbf{m} \}$ are the POVM effects used with the last batch.
Between optimisation rounds, if $\epsilon_E$ increases twice in a row, the POVM is reset back two iterations.
In addition, if there is an increase in $\epsilon_E$ for a total of ten times during a measurement procedure, the POVM optimisation is deemed converged and it is thus turned off, and the best POVM until that point is chosen for following measurements.
In AIM-ADAPT-VQE, we also modify the gradient-based POVM optimisation strategy.
Instead of having a defined gradient step size, after the gradient direction is estimated, a line search is done in order to find the step size that decreases the variance the most.
These modifications to the POVM optimisation procedure result in an improved performance over the method employed in Ref. \cite{Garc_a_P_rez_2021}.
Finally, further resource savings may be obtained by recycling the POVM between ADAPT iterations \cite{Korhonen2022}.
The assumption is that, if the state has not changed significantly between two consecutive iterations, the optimal POVM for the previous round may be close to the optimal POVM for the next round.
Therefore, we initialise the POVM based on this strategy.

\section{Results}
\label{sec:results}

\begin{figure*}[t!] 
    \includegraphics[width=\linewidth]{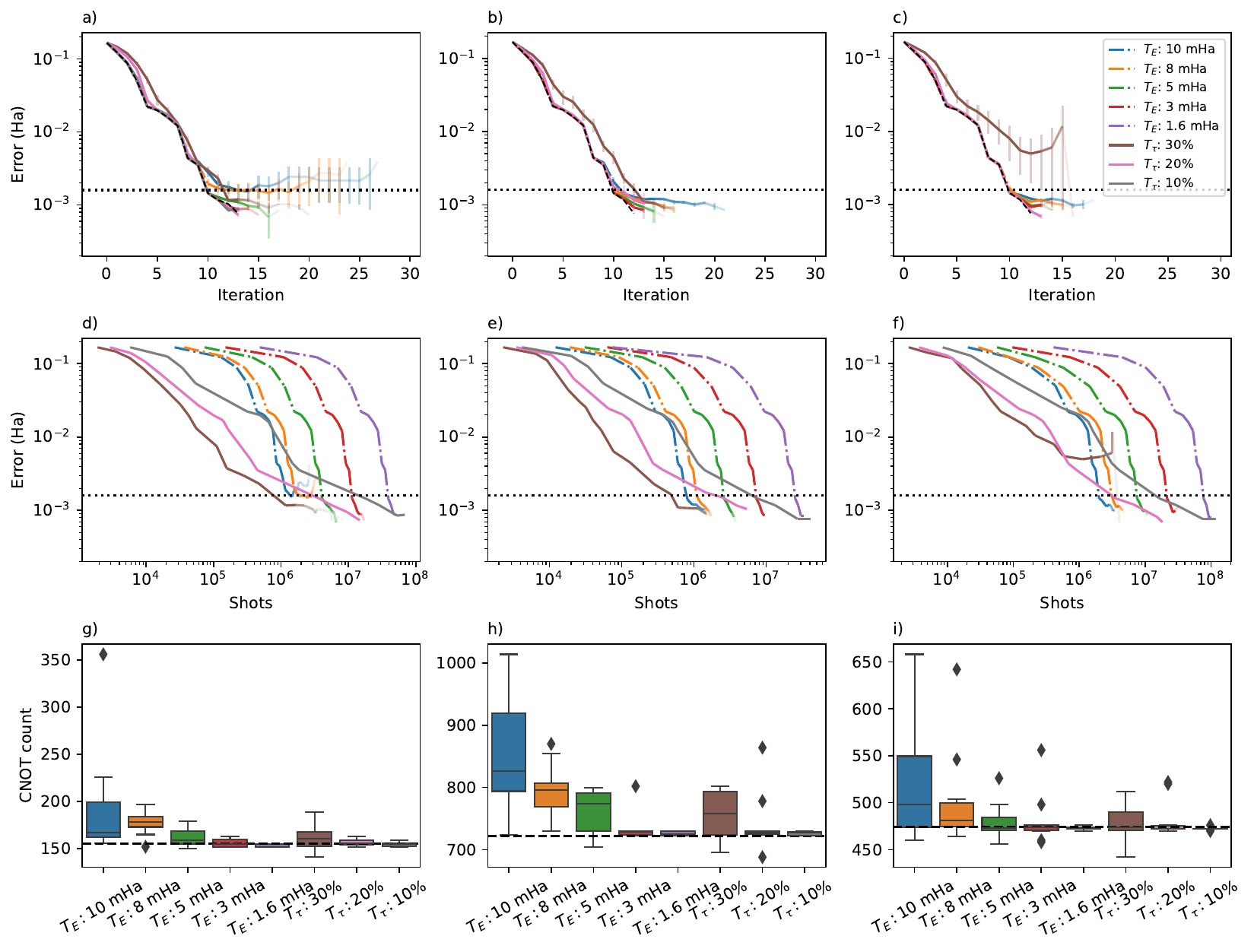}
    \caption{AIM-ADAPT-VQE simulations for different target measurement precisions for a hydrogen chain $\text{H}_4$ with interatomic distance $\SI{1.5}{\angstrom}$, using different fermion-to-qubit mappings and pools: a), d), g) JW mapping and QEB pool; b), e), h) BK mapping and spin-dependent fermionic pool; c), f), i) JKMN mapping and spin-dependent fermionic pool. All the results are obtained from 10 realisations of each experiment.
    The top panels, a), b), c), show the average error as a function of the ADAPT iteration.
    The dashed black line shows an exact statevector simulation and the horizontal dotted line the chemical precision ($\SI{1.6}{\milli\hartree}$). The error bars show one standard error distance from the average. The opacity of the lines is proportional to the number of runs that have not met the stopping criteria up until that point. 
    The second row of panels, d), e), f) show the average error as a function of the cumulative number of shots used by AIM-ADAPT-VQE. The dotted line corresponds to the chemical precision $\SI{1.6}{\milli\hartree}$. 
    The bottom panels, g), h), i), are box-and-whisker plots of the CNOT counts of the constructed circuits at the end of the AIM-ADAPT-VQE simulations. The boxes represent the distribution of CNOT counts over the realisations of each experiment. The dashed lines show the number of CNOTs obtained with a statevector simulation of ADAPT-VQE. 
    }
    \label{fig:fermionics}
\end{figure*}

In order to test AIM-ADAPT-VQE, we consider a 4-hydrogen chain ${\rm H_4}$ with bond distance $\SI{1.5}{\angstrom}$ and dinitrogen ${\rm N_2}$ with equilibrium bond distance $\SI{1.098}{\angstrom}$.
These are good case studies, as they are fairly complex molecules, with non-trivial correlations, so that ADAPT-VQE does not converge in too few iterations.

The fermionic Hamiltonians are obtained with the EOS-pyscf implementation in Algorithmiq's software framework \emph{Aurora} using STO-3G orbital basis set resulting in four and six active spatial orbitals, respectively.

The ${\rm H_4}$ fermionic Hamiltonian is then mapped into qubit space with JW, BK and JKMN fermion-to-qubit transformations, which results in three different 8-qubit Hamiltonians. For ${\rm N_2}$, only JW mapping is used, which results in a 12-qubit Hamiltonian.
For ${\rm H_4}$ with JW mapping, the simulations are run with QEB and qubit-ADAPT operator pools, and for BK and JKMN mappings, with spin-dependent fermionic and qubit-ADAPT pools. For ${\rm N_2}$, the spin-dependent fermionic pool is used.

Each spin-dependent fermionic and QEB gate with JW mapping is implemented with an efficient circuit from Ref.~\cite{Yordanov_2020} leading to a decrease in CNOT count.
Spin-dependent fermionic operators for BK and JKMN mappings are implemented with the operator evolution in \emph{Qiskit} \cite{Qiskit} with one Trotter step.
Each of the blocks in the Trotter approximation and the gates from qubit-ADAPT pool are implemented by \emph{Qiskit}, resulting in the usual exponential map circuit containing staircase-like sequences of CNOT gates.

As previously explained, in this work, we focus on the ADAPT-VQE operator selection cost.
Therefore, in order to avoid the computational cost of simulating the intermediate ansatz optimisation steps with shot noise, the ansatz is optimised with exact simulations where the state vector is used to compute expectation values (in the following, we will refer to this kind of simulations as ``statevector'' simulations for conciseness) using the L-BFGS-B optimiser \cite{Zhu1997}.
To improve the speed and precision of the gradient calculations, the analytic gradient method introduced in the Supplementary Information of Ref.~\cite{Grimsley2019} is used.
All ansatz parameters are optimised after each AIM-ADAPT-VQE iteration.
Once the ansatz parameters are optimised, we perform the IC measurements to determine the next operator to be added with AIM-ADAPT-VQE, and record the number of measurements needed.
The simulations presented in this paper have been obtained with the implementation of AIM-ADAPT-VQE in \emph{Aurora}.

\begin{figure*}[t!]
    \includegraphics[width=\linewidth]{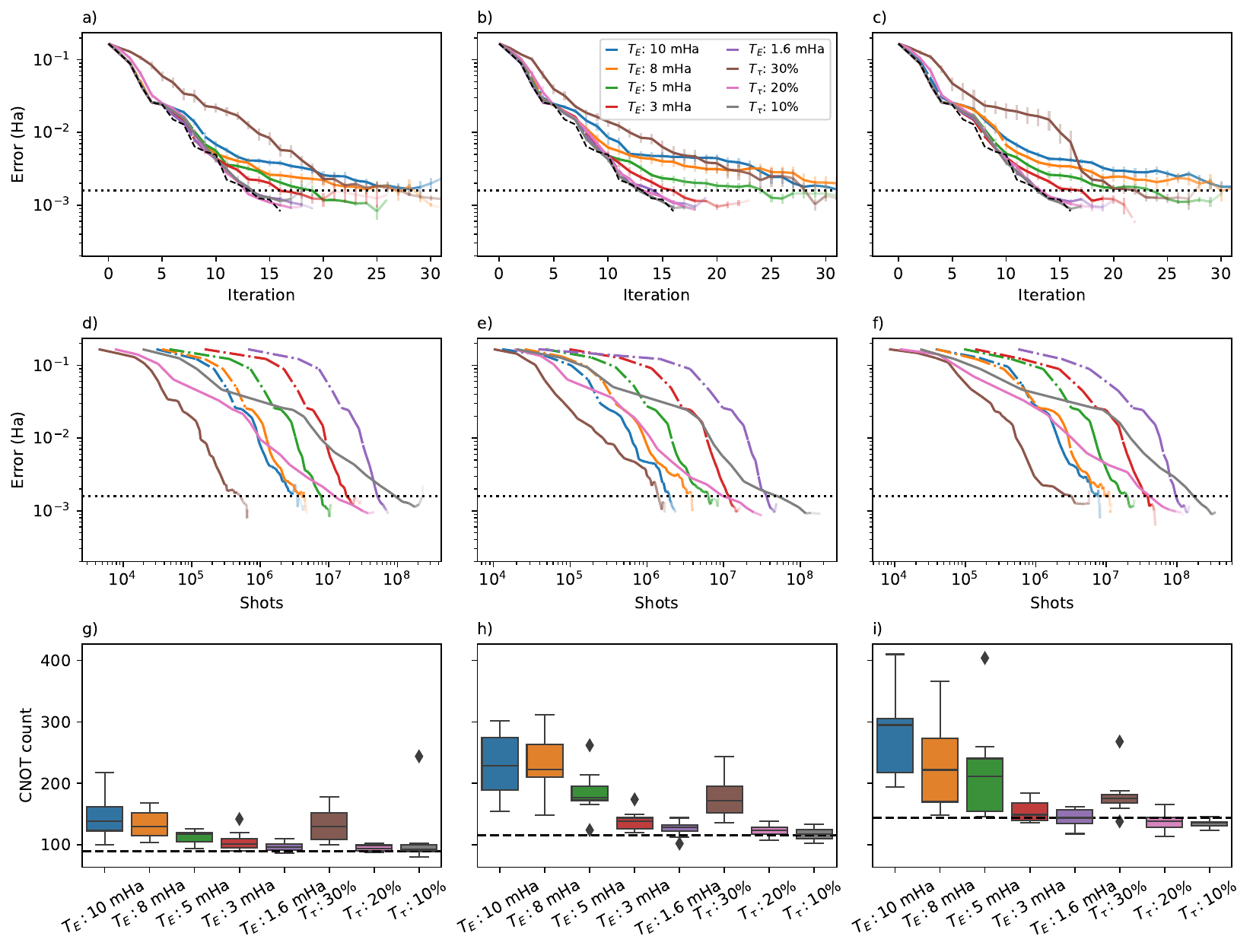}
    \caption{AIM-ADAPT-VQE simulations for $\text{H}_4$ with interatomic distance $\SI{1.5}{\angstrom}$, using the qubit-ADAPT pool and different fermion-to-qubit mappings: a), d), g) JW mapping; b), e), h) BK mapping; c), f), i) JKMN mapping. The same quantities as in Fig.~\ref{fig:fermionics} are shown. Top panels: average error vs number of iterations; mid panels: average error vs cumulative number of measurement shots; bottom panels: distribution of the CNOT counts of the transpiled final circuit.}
    \label{fig:qubit-adapt}
\end{figure*}

\subsubsection*{${\rm H_4}$ molecule}
 
In Fig.~\ref{fig:fermionics}a)--c), simulations of AIM-ADAPT-VQE for all three fermion-to-qubit mappings, QEB and spin-dependent fermionic operator pools and different convergence criteria, as well as target measurement precision, are shown.
Each line is an average over ten realisations, where each realisation is an AIM-ADAPT-VQE simulation run until the error to the exact ground state energy decreased below $\SI{1}{\milli\hartree}$ or the number of iterations reached 50.
The relation between the measurement precision and the rate of convergence can be clearly seen.
The simulations with higher precision measurements converge faster, while the most imprecise ones tend to require more iterations.

Importantly, the ground state is reached up to chemical precision by nearly all simulations.
In addition, by comparing them to the exact, infinite-statistics statevector simulation represented by the dashed black line, one can see that most of the realisations follow it quite closely.
In all cases, a threshold of $T_E = \SI{5}{\milli\hartree}$ or lower for energy-based target measurement precision, or $T_{\tau} = 20 \%$ or lower for relative error target precision in the gradient-based case, are required to converge in the same number of iterations as the exact simulation.

Measuring less precisely introduces randomness into the operator selection, making the runs behave differently, even amongst different simulations with the same target measurement precision.
This can be seen from the increasing opacity of some of the lines, which indicates that only a few runs fail to converge.
In Fig.~\ref{fig:fermionics}a), the $\SI{10}{\milli\hartree}$ and $\SI{8}{\milli\hartree}$ lines, as well as the $30 \%$ one in c), seem to suffer the most from this.

\begin{figure*}[t] 
    \includegraphics[width=\linewidth]{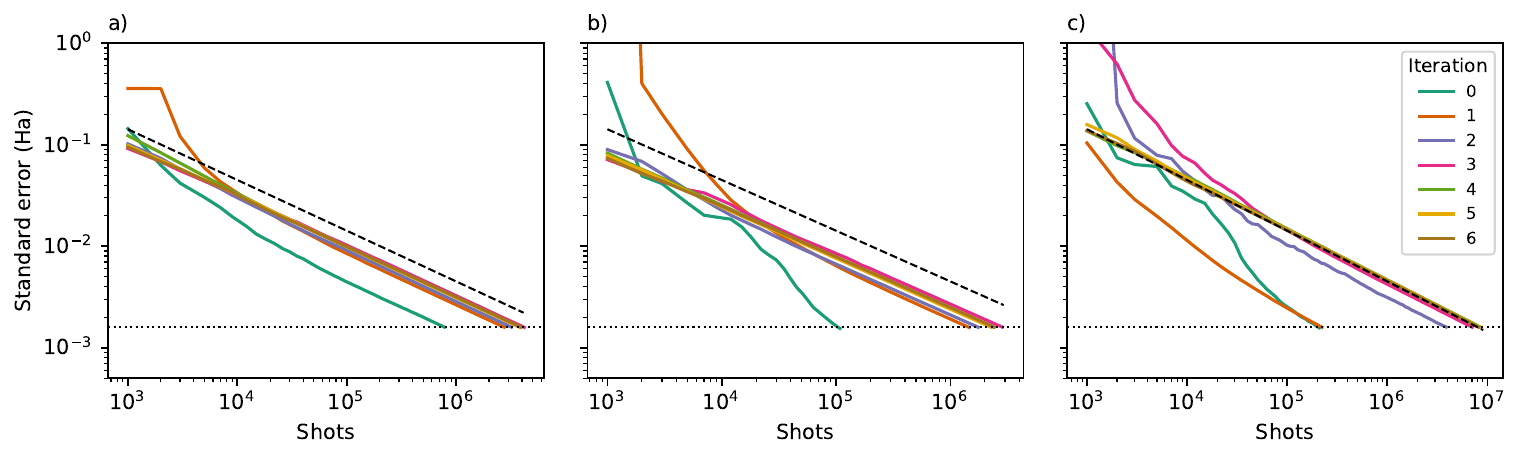}
    \caption{Optimisation of the IC POVM for different AIM-ADAPT-VQE iterations. The plots show the  estimated standard error in Ha as a function of the total number of shots used during steps 3-7 of the algorithm, for one run of the experiment. The panels correspond to those of Fig.~\ref{fig:fermionics}, for the strategy $T_E = \SI{1.6}{\milli\hartree}$. 
    The horizontal dotted line represents the chemical precision and the dashed line the shot-noise $1 / \sqrt{\text{shots}}$ scaling. At each iteration, the optimal POVM of the previous iteration is used as the initial condition. In the first iterations the lines decrease faster than the shot noise, meaning that the initial POVM is not optimal for the current ansatz and can be further refined by the algorithm. In later iterations the lines follow the shot-noise line, i.e., the POVM of the previous iteration is already quite close to optimal.}
    \label{fig:gradientoptimisers}
\end{figure*}

The same simulations with qubit-ADAPT operator pools can be seen in Fig.~\ref{fig:qubit-adapt}a)--c).
The effect of the target measurement precision on the rate of convergence is much larger here.
In all cases, an energy-based target measurement precision of $T_E = \SI{1.6}{\milli\hartree}$ or $T_{\tau} = 20 \%$ relative error of operator gradient target measurement precision suffice to converge as with an statevector simulation.
From all simulations with QEB, spin-dependent fermionic, and qubit-ADAPT operators pools, it can be concluded that the measurement scheme with relative error of operator gradient as target measurement precision seems to be a safer choice in ensuring that the convergence rate remains optimal.

The QEB and spin-dependent fermionic pools, which are chemically inspired, behave better with less precise measurements than the qubit-ADAPT pool.
As qubit-ADAPT pool is constructed by splitting a single excitation term into two or eight separate operators acting on the same qubits, one would expect significant redundancy in choosing the optimal gate and, therefore, that non-optimal choices would not affect the performance much.
Instead, as Fig.~\ref{fig:qubit-adapt} shows, this is clearly not the case.
A possible reason for this is that qubit-ADAPT pool can break the $S_z$ symmetry.
Non-optimal choices can introduce more symmetry breaking, leading the circuit easily into local minima.

An important metric to consider when constructing VQE ansätze is the number of gates required, as this affects both the execution time and, in near-term devices, the amount of noise introduced in the computation.
In particular, the number of CNOT gates is a relevant figure of merit as these are typically noisier than single-qubit gates.
In Fig.~\ref{fig:fermionics}g)--i), the resulting CNOT counts of each run are shown.
Each CNOT count is calculated after compiling the circuit with \emph{Qiskit}'s \verb|transpile| function with the highest optimisation level with all-to-all connectivity \cite{Qiskit}.
The performance can be evaluated by comparing the counts to the green line representing the CNOT count of an exact, statevector run.
In all cases, measuring until chemical precision is sufficient to reach the same CNOT count as the state vector simulation up to small deviations.

In many cases, measuring until chemical precision is not required, and one can match the statevector simulation result with less precise measurements.
However, measuring less precisely does introduce randomness and increases the probability of having a worse run.
The level of this effect differs for different fermion-to-qubit mappings.

Interestingly, many of the runs reach lower CNOT counts than the exact statevector simulation.
While the stochastic nature of the transpilation function employed may cause small deviations, this alone does not explain the result.
Instead, this shows that the gate ordering obtained from an ideal ADAPT-VQE algorithm with gradient-based selection might not be optimal, which is somewhat expected given that ADAPT-VQE is a greedy algorithm.
In any case, the improvement in the CNOT counts obtained here arise only from randomness in the operator selection due to imprecision in the gradient evaluations.

The analysis for different measurement precision levels allows us to examine how robust the AIM-ADAPT-VQE algorithm is to measurement data scarcity.
In Fig.~\ref{fig:fermionics}d)--f) (as well as in Fig.~\ref{fig:qubit-adapt}d)--f) for the qubit-ADAPT pool), the error in the energy as a function of the total number of measurements used is shown.
Importantly, it can be seen that, if the measurement threshold, either $T_E$ or $T_\tau$, is set to a high value, the ground state can be reached with much fewer shots than needed simply for energy evaluation within chemical precision (which is indicated by the $T_E = \SI{1.6}{\milli\hartree}$ curve).
However, this typically results in deeper circuits.
In any case, the gradient-based target measurement threshold seems to consistently perform better, since resources are spared in the first iterations, where the gradients are large and hence precise measurements are not needed to identify the largest ones.

It is interesting to discuss the effect of the initialisation of the POVM based on the optimality of the last round.
In Fig.~\ref{fig:gradientoptimisers}, we depict the error in the estimation of the energy as a function of the number of shots for different ADAPT iterations in which the POVM at each iteration is initialised to match the optimal previous one.
The results show that, after the first few iterations, the error does not decrease faster than the standard statistical inference curve, indicating that the optimisation of the POVM no longer improves its performance.
The erratic behaviour in the first iteration in all cases, as well as in the second and third iterations in Fig.~\ref{fig:gradientoptimisers}c), are caused by the fact that some of the qubits remain in the computational basis in the first iterations, and after optimising the POVM corresponding to those qubits, they perform very poorly in the next iteration when they are recycled to be used on entangled qubits.
In these cases, the POVM are reset.

The optimisation of the POVM only requires classical resources.
However, in the case when there is noise affecting the POVM implementation, quantum detector tomography can be used in order to reduce the bias induced by noise on the expectation values \cite{Glos2022,cattaneo2022semidefinite}.
When optimising the POVM, detector tomography needs to be performed each time the POVM is changed.
These results show that, since after certain number of ADAPT iterations the POVM stop benefiting from the optimisation, and hence the POVM optimisation can be stopped, the overall cost of detector tomography could be heavily reduced.

\begin{figure*}[t] 
    \includegraphics[width=\linewidth]{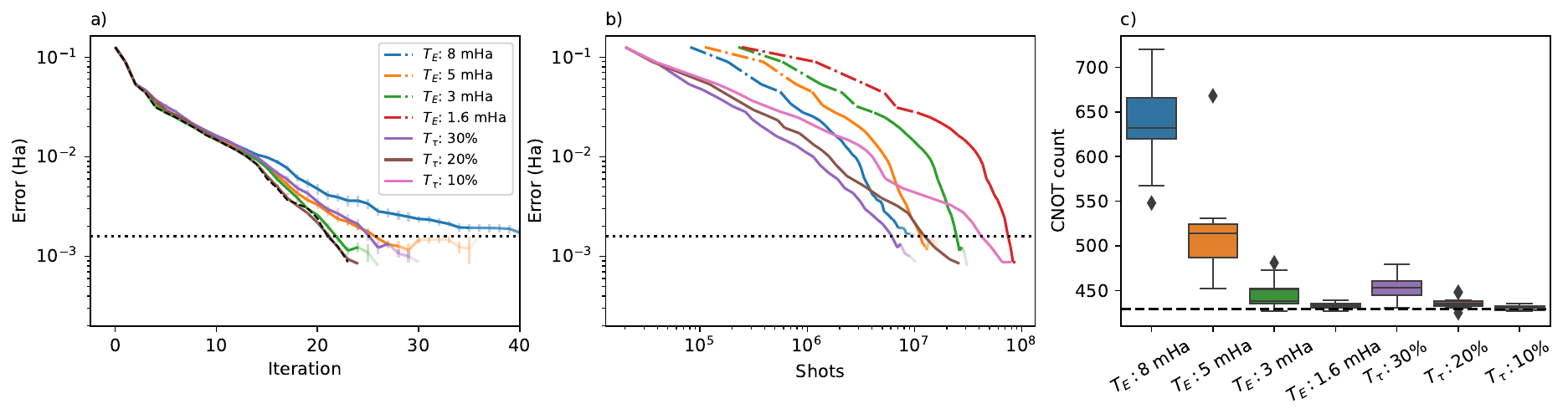}
    \caption{AIM-ADAPT-VQE simulations for $\text{N}_2$ with interatomic distance $\SI{1.098}{\angstrom}$, using the fermionic pool and Jordan-Wigner fermion-to-qubit mapping. The same quantities as in Fig.~\ref{fig:fermionics} are shown: a) average error vs number of iterations; b) average error vs cumulative number of measurement shots; c) distribution of the CNOT counts of the transpiled final circuit.}
    \label{fig:N2-fermionic}
\end{figure*}

\subsubsection*{${\rm N_2}$ molecule}

A similar analysis is performed for the larger example, ${\rm N_2}$, with the results shown in Fig.~\ref{fig:N2-fermionic}. Again, most of the simulations reach chemical precision within a number of iteration close to the statevector simulation, as seen in Fig.~\ref{fig:N2-fermionic}a) and c). A threshold of $T_E = \SI{3}{\milli\hartree}$ or lower for energy-based target measurement precision is required for minimal iteration overhead. The relative-error based target precision is again more efficient and $T_{\tau} = 20 \%$ or below is required to match the statevector simulation. The $T_{\tau} = 20 \%$ simulation convergences slightly faster than the  $T_E = \SI{3}{\milli\hartree}$ one, but requires less measurements as seen in Fig.~\ref{fig:N2-fermionic}b).

A critical point here is that the correct operator can be found without performing full state tomography on the state each iteration.
Full state tomography requires inferring all $4^N$ components in the system's density matrix, and therefore the number of measurements needed in state tomography is typically much larger than $4^N$.
In the case of a 12-qubit system like the one considered here, there are approximately $10^7$ elements in the system's density matrix.
Yet, from Fig.~\ref{fig:N2-fermionic}b), it can be seen that each ADAPT iteration requires much fewer shots than that.
While for ${\rm H_4}$ the measurement cost for the energy measurement is in the order of $4^N$, for ${\rm N_2}$ it is already considerably below it.
This indicates that as the system size increases, the required measurements cost for operator selection remains well below the tomographic regime, in agreement with theoretical expectations~\cite{Garc_a_P_rez_2021}.

\section{Conclusions and Outlook}
\label{sec:conclusions}

ADAPT-VQE algorithms provide an efficient toolbox to construct accurate and shallow circuits specific to the electronic structure problem.
Their major drawback is the measurement overhead arising from the gradient calculations that need to be performed in each iteration.
Here we have shown that this overhead can be mitigated, or even bypassed, by measuring with adaptive IC POVM.

By reusing the IC data obtained in energy measurement steps needed to optimise the ansatz, it is possible to estimate all the operator gradients with no additional measurement cost. Furthermore, for increasing system size, this selection can be done with a number of shots that is well below the tomographic regime.
For the examples studied in this work, the data collected for energy estimations within chemical precision sufficed to implement different versions of ADAPT-VQE without changing the CNOT counts of the original algorithm. In some cases the counts were even lower than with exact simulations.

In addition, we have shown that AIM-ADAPT-VQE is rather robust in terms of measurement precision and performs reasonably well even with less precise measurements. Some of the fermion-to-qubit mappings and operator pools considered were found to be more robust than others.
Interestingly, even when very few measurements are performed, the AIM-ADAPT-VQE routine eventually converges, albeit with higher CNOT counts.
This trade-off between measurement overhead and circuit complexity can be exploited to optimise the implementation of algorithms of the ADAPT family on slow but noise resilient hardware platforms, such as ion trap based systems.

The IC-based implementation of ADAPT algorithms offers other interesting and potentially useful ways to further optimise the ansatz construction.
For example, if IC data is available, it is possible to evaluate the energy of the system once a given gate is appended to the circuit using only classical post-processing.
This could open the path to a different operator selection strategy without additional measurement cost: instead of choosing the operators based on energy gradients, they could be chosen based on their impact on the energy once optimised.

Further, for AIM-ADAPT-VQE, the only computational cost that increases with the size of the pool is, by construction, classical (assuming similar algebraic properties in the pool operators).
This property motivates the search of larger operator pools that may be able to produce accurate ans\"{a}tze with even smaller CNOT counts. This approach is in stark contrast to most works related to operator pool selection, where the aim has been to reduce measurement costs by reducing the pool sizes without compromising performance.

The combination of ADAPT-VQE algorithms and adaptive informationally complete measurements provides a recipe for performing accurate quantum simulations with shallow circuits and reasonable measurement cost scaling.
We see this as an important step towards performing quantum simulations of systems with increasing system size, and therefore towards near-term quantum advantage for quantum chemistry.

\bibliography{references.bib}

\newpage
\appendix

\end{document}